\documentstyle[preprint,aps]{revtex}
\begin{document}
\draft
\title{\bf A note on the Einstein equation in string theory}
\author{Sayan Kar \thanks{Electronic Address :
sayan@iucaa.ernet.in}} 
\address{Inter University Centre for Astronomy and Astrophysics,\\
Post Bag 4, Ganeshkhind, Pune, 411 007, INDIA}
%\twocolumn[
\maketitle
%\widetext
\parshape=1 0.75in 5.5in
\begin{abstract}
We show, using purely classical considerations and logical
extrapolation of results belonging to point particle
theories, that the metric background field in which
a string propagates must satisfy an Einstein or an Einstein--like 
equation. Additionally, there emerge restrictions on the
worldsheet curvature, which seems to act as a source for spacetime
gravity, even in the absence of other matter fields.
\end{abstract}
\vskip 0.125 in
\parshape=1 0.75in 5.5in
%PACS number(s) :
%\pacs{}
%]

%\narrowtext 

\newpage

String theory, unlike General Relativity (GR),
is {\em not} a theory of spacetime. It is, ofcourse, not
expected to be so, primarily because the notion of a
string as a fundamental entity is the alternative 
to the usual point particle paradigm. Spacetime geometry makes its
appearance in
string theory in a very different way.  

Firstly, quantum string theory is believed to include a quantum theory of
gravity apart from unifying all the forces of nature. 
Furthermore, the low energy, effective field theory that emerges out of
string theory resembles General Relativity (GR) in the Einstein frame
(and a Brans--Dicke (BD) type
theory in the string frame) {\cite{gsw:book}}. This remarkable fact leads us to 
believe that classical spacetime geometry 
emerges in the low--energy limit of string theory. The main cause
behind it is the assumption that quantum conformal
invariance must hold good even if one breaks classical
conformal invariance by including the dilaton dependent term in the
$\sigma$ model action. This enables us to set the $\beta$
function(al)s of the $\sigma$--model action to zero. The resultant partial
differential equations
constitute a set of constraints on the metric, axion (or antisymmetric
tensor field) and dilaton 
fields which, actually are, couplings at the level of the
$ \sigma$ model. These constraints on the couplings,
surprisingly, turn out
to be the above--mentioned Einstein--like equations and constitute the
low energy effective field equations for full string theory. 
  
Is there a purely classical way through which one can prove that the
conditions on the background metric/matter fields (couplings)
 in the $\sigma$--model action
must resemble the Einstein equations? We shall outline in this
paper, a possible path through which this can be achieved. It is based on the
extrapolation of the well--known relationships between the geodesic 
deviation equation, the geodesic
equation and the Einstein field equation to the case of extended objects,
or more specifically, strings.
The logic proceeds as follows.

We recall the following facts from the theory of point particles and
General Relativity (GR). 

(i)The geodesic deviation equation is generically of the form,
\begin{equation}
\frac{D^{2}{\eta}^{i}}{D\tau^{2}} + K^{i}_{j} \eta^{j} = 0
\end{equation}

where $\frac{D}{D\tau}$ is the covariant derivative along the
tangent vector to the world--line(geodesic). $\eta^{i}$ is the
deviation vector normal to the geodesic line. The indices
$i, j$ are associated with the normal--frame attached at each
point on the geodesic--therefore, $i=1,2,3$ for a point particle
trajectory in four dimensions . 

This equation is true for Newtonian theory as well. We shall show, later in 
this article,
that the condition $Trace (K^{i}_{j}) = 0$ yields the Laplace equation
for Newtonian theory (i.e. Poisson's equation in the absence of
matter)  and the vacuum Einstein field equations for
a relativistic theory in curved spacetime. This will be
one of our guiding principles in the attempt to arrive
at an equation for string background fields in a classical
framework.   

(ii) Einstein's equation automatically implies a conservation law. If we add
a test--particle action to the Einstein--Hilbert + other matter terms
 and obtain
the corresponding spacetime energy--momentum tensor due to the
test particle, its conservation law straightaway leads to the
geodesic equation. Thus the field equation contains
the equation for test particles--a fact which is not true for
other field theories such as electrodynamics (the Lorentz
force equation {\em cannot} be derived in any way from
the Maxwell equations). This unique feature
of Einstein's theory is attributed partly 
to its non--linearity {\cite{eg:pap}. 

We shall extrapolate both these statements to the case of strings
and investigate the corresponding consequences. The end result will reveal
 that
the compatibility conditions on the metric tensor does lead to
the Einstein equations and the conservation law for the test
string spacetime energy--momentum tensor implies the
string equation of motion in a general curved background.
Therefore, it is {\em no surprise} that the background metric
field for string propagation should satisfy an Einstein--like
equation. Moreover, the entire analysis of this paper will be based
on classical physics which is quite different from the usual
way in which string theory gives rise to GR (or Brans--Dicke type
theories). 
 
Let us begin by recalling the Newtonian deviation equation (a clear derivation
of deviation equations for point particle theories 
can be found in {\cite{rdi:book}} ). 
If $\eta^{i}$ is the deviation vector and $\phi$ the
Newtonian gravitational potential, we can easily show that the 
equation is :

\begin{equation}
{\ddot\eta}^{i} + \left ( \partial^{i}\partial_{j}\phi \right ) \eta^{j} = 0
\end{equation}

where a dot denotes a derivative with respect to time $t$.
Thus we get, 

\begin{equation}
Trace (K ^{i}_{j}) = Trace (\partial^{i}\partial_{j}\phi) = 0 \qquad \Rightarrow \qquad 
\nabla^{2} \phi = 0
\end{equation}

which is the vacuum (Laplace) equation for the gravitational potential $\phi$.

In a curved spacetime, the deviation equation remains of a similar form
(it is as stated in (1) ),
but the quantity $K^{i}_{j}$ turns out to be :

\begin{equation}
K^{i}_{j} = R_{\mu\nu\rho\sigma}E^{\mu}E^{\rho}n^{\nu}_{j}n^{\sigma i}
\end{equation}

where $E^{\mu}$ and $n^{\mu}_{i}$ are the tangent and normals to the
geodesic curve.

Once again, it is fairly easy to see that :

\begin{equation}  
Trace (K ^{i}_{j}) = 0 \qquad \Rightarrow \qquad 
R_{\mu\nu} = 0
\end{equation}

where we use $n^{\mu i}n^{\nu}_{i} = g^{\mu\nu} + E^{\mu}E^{\nu}$. 
Thus, we arrive at the vacuum field equation of GR by utilising the
same principle.

To extend the above logic to the case of string worldsheets we need to
know the deviation equation for string worldsheets of Nambu--Goto
/Polyakov type. This, fortunately,
has been there in the literature for quite some time now {\cite{jac:many}}
. It is related, as in the geodesic case, to the second variation
of the Nambu--Goto action evaluated at its stationary points and reads :

\begin{equation}
\Box \eta ^{i} + \left ( M^{2}\right )^{i}_{j} \eta^{j} = 0
\end{equation}

where $\Box$ denotes the usual D'Alembertian operator on the string
world sheet. The $\eta^{i}$ are the deviation vector components
in the normal directions. More specifically if $\delta x^{\mu}$ is the
deformation of the embedding function we have 
$\delta x^{\mu} = n^{\mu}_{i}\eta^{i} + E^{\mu}_{a}\eta^{a}$ where  
$E^{\mu}_{a}$ are the tangent vectors to the string world sheet with
$a=\sigma, \tau$. We ignore  
the tangential deformation $\eta^{a}$, because, by reparametrisation
invariance we can say that this does not cause any changes in the worldsheet 
geometry.  
   The quantity
$\left (M^{2}\right )^{i}_{j}$ (which replaces the $K^{i}_{j}$
mentioned earlier)  is given as :

\begin{equation}
\left ( M^{2} \right )^{i}_{j} = K^{ab i}K_{ab j} + R_{\mu\nu\rho\sigma}
E^{\mu}_{a}n^{\nu i}E^{\rho a}n^{\sigma}_{j} 
\end{equation}

where $K^{ab i}$ denotes the extrinsic curvature of the embedded string
world--sheet in the direction of the $i$th normal $n^{\mu i}$. The 
$E^{\mu}_{a}$ are the tangent vectors to the string world sheet with
$a=\sigma, \tau$. Note that the $\left ( M^{2}\right )^{i}_{j}$ 
by definition is dependent on the choice of the normal frame.
However, when we take a trace this ambiguity goes away and we
get a quantity which is independent of the choice of normals.  
(The same is true for the case of geodesic curves). It should also
be stated that we have implicitly assumed in our deviation equation the
existence of 
 those extremal worldsheets
for which the normal fundamental form ($\mu_{ij}^{a}=g_{\mu\nu}n^{\mu}_{i}
E^{\alpha a}D_{\alpha}n^{\nu}_{j}$) vanishes. This is not too restrictive
as can be seen by inspecting $\mu_{ij}^{a}$ for the multitude of string
configurations available in various curved background geometries. 

Therefore, on extrapolating the condition for the point particle
case to strings , we get :

\begin{equation}
Trace(K^{i}_{j}) = Trace[(M^{2})^{i}_{j}] =
 -^{2}R + R_{\mu\nu}E^{\mu a}E^{\nu}_{a} = 0
\end{equation}

where we have used the traced Gauss--Codazzi integrability condition
and the extremality condition $K^{i} = 0$ in order to arrive at
the above expression after taking the trace with respect to the
normal indices.

The above criterion has been obtained in a different way recently
in {\cite{pv:prd97}}. There, once again, one utilises the requirement 
of one--loop finiteness of Nambu--Goto string theory.
The authors in {\cite{pv:prd97}} propose this as a consistent
condition for one--loop finiteness, by writing the above equation as
an equation on the worldsheet. We shall show that the condition
can be split into two separate conditions--one on the worldsheet and one on
spacetime geometry in a very simple way.
  
It should be mentioned that the quantity $Trace(K^{i}_{j})$ appears
in the generalised Raychaudhuri equation for worldsheet congruences
of Nambu--Goto strings {\cite{grchd:prd95}} which is given as :

\begin{equation}
\Box F + \frac{1}{N-2}\left ( -^{2}R + R_{\mu\nu}E^{\mu a}E^{\nu}_{a} \right ) F  = 0
\end{equation}

where $\theta_{a} = \frac{\partial_{a}F}{F}$ are the expansions along
the $\sigma$ and $\tau$ directions on the worldsheet.
This fact is true for the Raychaudhuri equation for geodesic congruences
in Newtonian as well as Riemannian spacetimes. 

Also note that in both the deviation equation for
extremal worldsheets and the generalised Raychaudhuri equation 
there appears contributions from the worldsheet geometry--the
$K_{ab}^{i}K^{ab}_{j}$ term in the deviation equation and the
$-^{2}R$ term in the generalised Raychaudhuri equation. Thus
even in the absence of spacetime gravity, worldsheet extrinsic
curvature can cause them to deviate and worldsheet intrinsic
curvature can lead to worldsheet focusing effects. This is,
however, expected to happen for extended objects and should not
be considered as too surprising !
  
What is the corresponding vacuum Einstein equation ?
Notice that the first term in the above expression (i.e. $^{2}R$) in 
$Trace(K^{i}_{j})$  is an explicit function of
$\sigma,\tau$ while the second term is a mixture of terms
which are explicit functions of $x^{\mu}$ ($R_{\mu\nu}$)
 as well as terms which are
explicit functions of $\sigma, \tau$ ( the tangent vectors
$E^{\mu}_{a}$ ). The only way in which we can have $Trace (K^{i}_{j}) = 0$
is to equate each term to the same quantity. Therefore, specifically, we
can have,
 
\begin{eqnarray}
R_{\mu\nu} = \Lambda \left (x \right ) g_{\mu\nu} \\
^{2}R = 2\Lambda \left (x(\sigma, \tau ) \right )
\end{eqnarray}

The first of these equations is, in general, in spacetime. One can
also evaluate it on the worldsheet by using the embedding
function $x^{\mu}(\sigma,\tau)$.
The second equation, on the other hand,
 is exclusively on the worldsheet itself and
one cannot convert it into a spacetime equation because
the embedding functions cannot be inverted.
To satisfy a conservation law, one needs $\Lambda$ to be a constant.
Thus, vacuum spacetimes in which string worldsheets can propagate
are essentially Einstein spaces {\cite{eis:riem}}. Note that this also includes
spacetimes with $R_{\mu\nu} =0$ (the $\Lambda =0$ case).
Worldsheet curvature seems to act as
a cosmological constant. Thus the existence of string worldsheets
with a nonzero worldsheet curvature essentially imply the existence
of a finite cosmological constant. 

Equating the $Trace(K^{i}_{j})
$ to zero in Newtonian theory and in GR amounts to considering
 $vacuum$ spacetimes. 
However, vacuum, in the presence of strings, as opposed to point particles
naturally contains worldsheet curvature which can act as a source for
spacetime gravity. This is an important difference which one must
remember while dealing with extended objects. Therefore, the
vacuum equation in the presence of strings is {\em not} just
the equation $R_{\mu\nu} = 0$ which appears as a special
case when we consider only flat worldsheets. For $p$-- branes the
same rule applies -- the only difference is that $^{2}R$ is
now replaced by $^{p+1}R$ -- the Ricci scalar for the $p$--brane
world--surface.

We now illustrate the above facts with an example.
The simplest possiblity is to look at spacetimes which
satisfy $R_{\mu\nu} = 0$. Correspondingly, one would require
$^{2}R = 0$. Minkowski spacetime is a trivial example. For
instance if we consider Minkowski spacetime in spherical
polar coordinates, the closed string solution
$t=\tau$, $r=r_{0}$, $\theta = \frac{\pi}{2}$ and 
$\phi = \sigma$ is an admissible string worldsheet.
The induced world--sheet metric here is consequently flat
Minkowski space in two dimensions and thereby has a zero
value for the Ricci scalar.

For worldsheets in $2+1$ dimensions it is easy to see that the
extremality condition $K=  \frac{1}{2} (K_{1} + K_{2}) = 0$ 
(where $K_{1,2}$ are the
principal curvatures) clearly implies $^{2}R = K_{1}K_{2}= -K_{1}^{2}$
 be a non--positive 
quantity. Therefore, the background spacetime could be a geometry
with a negative cosmological constant with its value 
being equal to the negative of the square of one
of the principal curvatures of the embedded worldsheet.
Thus, the black hole spacetime of Banados, Teitelboim and
Zanelli {\cite{btz:prl92}} can be an admissible background geometry. 
However, one has to make sure that constant negative curvature  
worldsheets exist in such backgrounds.

More generally, it is possible that one equates the
 $R_{\mu\nu}$ to an object related to 
the energy--momentum tensor for an extra, auxiliary field $\chi$. In a
somewhat general setting one may write the resultant Einstein
equation as :

\begin{equation}
G_{\mu\nu} = \Lambda \left (\chi\right ) T^{\chi}_{\mu\nu}
\end{equation}

and the corresponding expression for $^{2}R$ as :

\begin{equation}
^{2} R = \Lambda \left ( \chi \right ) \left [ \frac{N-4}{N-2} T^{\chi}
 - T^{\chi i}_{i}
\right ] 
\end{equation}

where $T = g_{\mu\nu} T^{\mu\nu} = T^{a}_{a} + T^{i}_{i}$ using
$g_{\mu\nu} = E^{\mu}_{a}E^{\nu a} + n^{\mu i}n^{\mu}_{i}$.

We therefore have a Brans--Dicke type theory where the worldsheet
curvature is partly responsible (through its trace and normal/
tangential projections) for generating the energy momentum
tensor for the Brans--Dicke field and hence the variable 
,effective $G$. This is, in some way,  absolutely necessary
if we intend to include worldsheets of non--constant curvature, which,
in actuality, is the most general situation. It is obvious to
ask, why does a {\em vacuum} Brans--Dicke type theory emerge in this process?
 The answer 
is as follows. If one allows for a constant $\Lambda$ in the 
equation with a $T_{\mu\nu}$ then one is assuming a priori extra
matter other than that which can generate a variable $G$.
Therefore, the equation, which is supposed to be a $vacuum$
equation becomes somewhat ambiguous. If, on the other hand,
, we have a BD type scalar field, then we can argue that
we still have a vacuum equation but the effective $G$ is now
generated through this scalar field, which, in turn, is
partly related to the worldsheet curvature.   

We have therefore demonstrated via the extrapolation of the
result (i) quoted in the beginning of the paper to the case
of strings that the background field  $g_{\mu\nu}$ must satisfy an
Einstein equation. Additionally, we find that the Ricci
scalar of the worldsheet is related in some way to the 
R. H. S. of the Einstein equation. We therefore notice
that the existence of a worldsheet configuration is
intimately coupled to the nature of the background 
geometry apart from its being an extremal configuration
with zero mean curvature along the normal directions.   
 
Let us now see what happens if we extrapolate the result in
(ii). To this end, we need to write down the expression for
the spacetime energy--momentum tensor of a test--string. 
This is obtained by simply varying the Polyakov action
with respect to the target space metric $g_{\mu\nu}$.
We have,

\begin{equation}
\sqrt {-g} T^{\mu\nu} = \frac{1}{2\pi\alpha^{\prime}}
\int d\tau d\sigma \left ( {\dot x}^{\mu}{\dot x}^{\nu}
- x^{\mu\prime}x^{\nu \prime} \right ) \delta^{(D)} \left ( x - x(\tau, \sigma)
\right ) 
\end{equation}

where we have chosen to work in the conformal gauge.

The $\delta$ function in the above expression ensures that we
are evaluating the $T^{\mu\nu}$ on the worldsheet. Note also
that in the above expression there is no assumption about the
extremality ($K^{i} = 0$ ) of the worldsheet. The test--string
action when added to the gravity + matter action is assumed not
to effect the structure of the background geometry..
We now impose the conservation law $T^{\mu\nu}_{;\nu} = 0$ and
see if we get back the string equation of motion. The steps are
outlined below {\footnote { A general derivation of this fact
 for actions
containing Nambu--Goto + rigidity corrections can be found in
{\cite{pv:prd97}}}}.
  
Evaluating the covariant divergence of the $T^{\mu\nu}$ amounts to
the calculation of the covariant divergence of the quantity within
brackets in the integrand. We therefore have :

\begin{equation}
\nabla_{\mu}\left ( e^{\mu}_{\tau}e^{\nu}_{\tau} - e^{\mu}_{\sigma}
e^{\nu}_{\sigma} \right )= 0
\end{equation}

where we have denoted $e^{\mu}_{\tau} = {\dot x}^{\mu}$ and 
$e^{\mu}_{\sigma} = {x}^{\prime \mu}$.

On explicitly writing out the action of the covariant derivative,
we obtain, 

\begin{equation}
e^{\mu}_{\tau}\nabla_{\mu}e^{\nu}_{\tau}+ 
e^{\nu}_{\tau}\nabla_{\mu}e^{\mu}_{\tau}- 
e^{\mu}_{\sigma}\nabla_{\mu}e^{\nu}_{\sigma}- 
e^{\nu}_{\sigma}\nabla_{\mu}e^{\mu}_{\sigma} = 0
\end{equation}

We now contract this equation first with $e_{\nu\tau}$.
This yields, on using the fact that $e^{\mu}_{a}e_{\mu b} = \Omega^{2}(\xi)
\eta_{ab}$ and $e_{\nu a}\nabla_{\mu}e^{\nu}_{a} = 0$ ( $a$ not summed here), 

\begin{equation}
\Omega^{2} \nabla_{\mu}e^{\mu}_{\tau} = 0
\end{equation}

Similarly, on contracting with $e_{\nu \sigma}$ we obtain,

\begin{equation}
\Omega^{2} \nabla_{\mu}e^{\mu}_{\sigma} = 0
\end{equation}

Substituting both these expressions into the covariant derivative
of the quantity in brackets in the integrand of the spacetime
energy--momentum tensor we obtain,

\begin{equation}
{\ddot x}^{\mu} - {x^{\mu ''}} + \Gamma^{\mu}_{\rho\sigma } \left
( {\dot x}^{\rho} {\dot x}^{\sigma} - {x}^{\rho \prime} {x}^{\sigma \prime}
\right )  = 0
\end{equation}

This is the string equation of motion in a generic curved spacetime!
Therefore, from the covariant conservation law of the spacetime
energy--momentum of the test--string we have arrived at the
string equation of motion. 

Thus, by a simple extrapolation of known facts from point particle
theories we have been able to derive the following results :

\begin{itemize}
\item{The compatibility conditions on the metric tensor which
appears as a coupling in the $\sigma$ model action turn out to
be the vacuum Einstein equation/the condition for a
spacetime to be an Einstein spacetime or the vacuum 
Brans--Dicke field equation.
This conclusion is based on the extrapolation of the relation 
between the geodesic deviation equation and the vacuum Einstein equation
to the case of string world--sheets.}
\item{The conservation law for the spacetime energy momentum tensor
of a test--string leads to the string equation of motion in
a generic curved background. Thus, the field equations contain the
equation for test--strings (similar to the point particle case) through
the conservation law.
But a field equation governing the dynamics of the background gravitational 
field can only  
be the Einstein equations/Brans--Dicke field equations if we demand  
an automatic satisfaction of the conservation
law through the Bianchi identity ! Note that our conclusion is dependent
on the assumption that the equation for test--strings be contained
in the field equation, which is an extrapolation of the corresponding
fact in the point particle case.  }
\end{itemize}

We believe that these results can be generalised to other $\sigma$
models as well, which include the axion, dilaton fields or
possess world--sheet/spacetime supersymmetry. 

The crucial feature about the above `derivation' is the fact that
we have exclusively used results of {\em classical} physics.  
Therefore, if we have strings as opposed to point particles
the background spacetime geometry should essentially obey the
same Einstein/Brans--Dicke type field equations with the curvature
of the worldsheet in some sense acting as the source for
spacetime gravity even in the absence of other sources of matter stress
energy.   

The author thanks S. Bose, S. Das, S. K. Rama, K. Ray, A. Sen, and Y. Shtanov 
for comments and  discussions. Financial support from the Inter--University
Centre for Astronomy and Astrophysics is also gratefully
acknowledged.

\end{document}